# Home Is Where the Up-Votes Are: Behavior Changes in Response to Feedback in Social Media[1]

SANMAY DAS and ALLEN LAVOIE, Washington University in St. Louis

## 1. INTRODUCTION

With the rise of peer production and social media as prominent creators and aggregators of information, questions of incentives and motivation are becoming more central. Why do users contribute? How do they decide where to contribute? There is an emerging consensus that social feedback is a major factor driving online behavior. Badges are used by venues to incentivize certain behaviors [Anderson et al. 2013]. Social feedback increases participation on YouTube and Digg [Wu et al. 2009], and on Wikipedia [Zhu et al. 2013]. It is clear that social feedback influences the decision to act or not act in many online settings, but its influence on more complex decision making is less certain.

Consider an example: A user posts a comment in a particular online community, receiving comment replies and a voting score from other users. All else being equal, the user is more likely to visit this community in the future. How much more likely? What is the exchange rate between positive ratings and comment replies? If the user gets more comment replies but fewer positive ratings in another community, how will she divide her efforts between these communities?

We introduce a model of human behavior changes in response to distributed—user-to-user—social interactions. This model provides explicit quantitative predictions for user behavior. In doing so, it creates a framework for evaluating the relative importance of different types of social feedback in determining user behavior, a form of inverse reinforcement learning. We evaluate the model's predictions on a community selection dataset collected from the social media website reddit.com.

As a model-free motivating example from this dataset, Figure 1 illustrates the effect of a particular kind of social feedback, comment replies, on the behavior of reddit users. Those who receive more comment replies in a community ("subreddit") are more likely to participate in that community in the future. There is a clear learning effect, with users spending more effort in communities where they have received more social feedback in the past.

Quantifying the motivating effects of social feedback allows us to predict individual user behavior, and as a consequence makes predictions for collective behavior. Users give feedback in communities they choose to participate in, and their choice of community is then influenced by the feedback they receive. The resulting collective dynamics shape user migrations between online communities.

## 2. MODELING COMMUNITY SELECTION

Social feedback can increase a user's propensity to perform an action: it can tip the scales when the choice is between doing and not doing. What if the choice is more complex? For example, a user visits reddit.com. Reddit is partitioned into communities, and the user must decide, implicitly or explicitly, which community to participate in. How does social feedback affect this choice? Given the visit, there

---

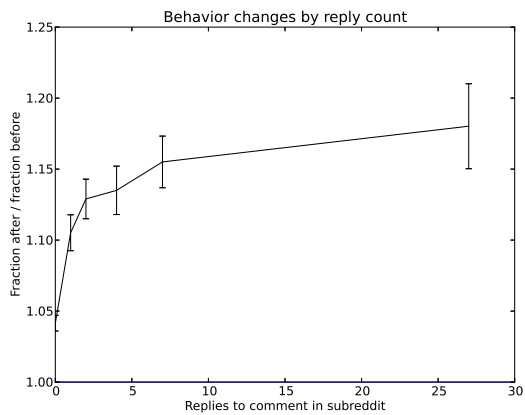

Fig. 1: Users change behavior in response to social feedback. In this model-free view, we compute the relative increase in visit frequency following social feedback (comment replies) in a community. Users who receive more comment replies are more likely to visit the community where those replies occurred in the future, consistent with a learning effect in response to social feedback.

**ALGORITHM 1:** Generative model of a user's behavior, based on initial propensities and learning in response to social feedback. $q$ is a vector of propensities, $q^0$ a user's initial propensities, and $s_i$ a chosen community. Selection is governed by Hierarchical Dirichlet Process parameters $\alpha_0$ (scalar concentration parameter) and $\beta$ (global community popularity vector), and by learning parameters $\phi$ (recency) and $\epsilon$ (exploration). A reward function $R$ determines the relative importance of social feedback features received in response to an action, which are encoded in vector $r_i$ (assumed to be instantaneous for simplicity).

$q^0 \sim \text{Dirichlet}(\alpha_0\beta)$ // Initial propensities (HDP)
$q \leftarrow q^0$;
**for** $i \in C_u$ **do** ;                    // User's actions (ordered)

$\quad s_i \sim \text{Categorical}(q/\sum_j q_j)$ ; // Strategy picking
$\quad q \leftarrow q(1 - \phi)$ ;                              // Recency
$\quad q_{s_i} \leftarrow (1 - \epsilon)R(r_i) + q_{s_i}$ ;       // Direct reward
$\quad q \leftarrow q + \epsilon R(r_i)q^0$ ;                   // Exploration
**end**

is an inherent competition between communities for that user's attention.[2] We model this conflict by analogy to human game playing. A player picks a strategy, and receives some reward based not only on his own action, but also the actions of other players. Picking a community is equivalent to picking a strategy, and the resulting social feedback is indicative of a reward.

We begin with the model of Erev and Roth [1998]: users maintain propensities, updated whenever a reward is received. The model is inherently probabilistic, with users drawing a strategy from a distribution defined by those propensities. With additional considerations for recency and exploration, this model has been quite successful in explaining and predicting human game playing behavior. However, the analogy between game playing and community selection is not perfect.

One important consideration is a user's exogenous interests (hobbies, physical location, etc.), each more or less popular globally (e.g. more people live in New York than St. Louis). While there are more online communities than there are strategies in the typical behavioral economics experiment, a more fundamental difference is that online communities are not finite in number: new communities are created frequently. The Hierarchical Dirichlet Process [Teh et al. 2006] captures these features: an infinite space of communities, individual preferences, and global popularities.

How motivating is a particular kind of social feedback? We employ a form of inverse reinforcement learning [Ng and Russell 2000]. Rather than assuming that an agent acts optimally with respect to an unknown utility function, we assume that agents learn according to the Erev and Roth model with rewards having an unknown relationship to social feedback. This learned relationship between social feedback and numeric reward, which we assume to be linear, is built into the model.

Combining these considerations, Algorithm 1 presents a generative model of user behavior in community selection. Users draw initial propensities according to the Hierarchical Dirichlet Process, then make some number of contributions. For each contribution, the user first picks a strategy (i.e. a community) by normalizing her propensities. The user then updates her propensities according to the reward received, which is a linear function of the observed social feedback (in our data, $r_i$ is a two-element vector containing a normalized reply count and voting score). Recency implies that older experiences are less influential, and exploration prevents a user from converging on a single community. Having specified a generative model of human behavior, we perform approximate Bayesian inference to learn the parameters of the model (recency, exploration, the reinforcement function, HDP parameters) and subsequently make predictions.

---

[2]Our analysis assumes the user has decided exogenously to participate or not, e.g. in reddit; this assumption could be relaxed by modeling a user's overall participation rate or adding competition with a "life" action [Anderson et al. 2013].

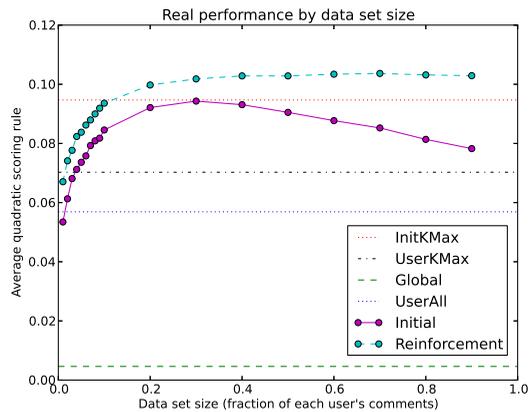

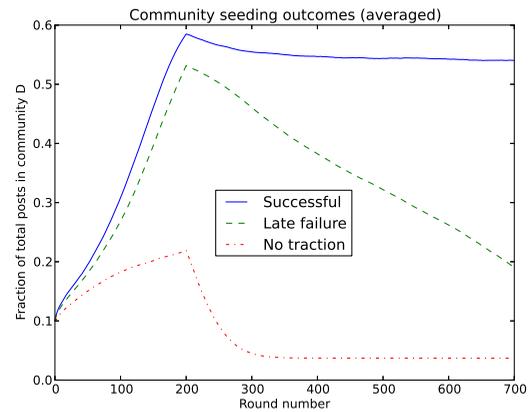

Fig. 2: Model performance on a probabilistic prediction task (community selection), as the fraction of each user's actions available for training is varied (with a consistent test set). Static models such as "Initial" (the generative model with no learning) become bogged down with irrelevant older data; this decline can be prevented with recency heuristics (e.g. InitKMax), but is further indication that a dynamic model is more appropriate. Other baselines are based on a model-free view of a user's previous actions (UserAll and UserKMax) or all users' actions (Global). The full reinforcement learning model performs significantly better than these baselines.

Fig. 3: Outcomes of community seeding simulations, modeling agents with parameters inferred from real data. Normal users select a community, then provide some feedback within that community (via voting and commenting). A set of seed users select and provide feedback only in a particular community for 200 rounds, their goal being to make that community dominant and self-sustaining. Sequences were grouped based first on the fraction of interest at round 200: no traction ($\leq 0.4$) or early traction. Of the latter, there are late failures ($\leq 0.5$ at 700) and successfully seeded communities. Curves are averaged within each group.

## 3. RESULTS

Figure 2 shows an evaluation of the predictions of the generative model on our reddit community selection dataset, containing 174783 submissions and comments by 1696 users. For each submission or comment, we collected its voting score and any replies in order to compute social feedback features. These predictions are probabilistic, evaluated with the quadratic scoring rule on observed outcomes. Our model performs significantly better than several plausible baselines.

We now turn to simulations of collective dynamics. How does a new social media site start? Without users, there is no content, and without content a site is not attractive to new users. Reddit itself famously began by presenting fake content posted by fake users [Morris 2012], but quickly became self-sustaining. Is a similar feat possible with social feedback? Figure 3 presents simulations in which a new community, initially very small, attracts new users by providing manufactured social feedback for a short period. Under the model of individual behavior we infer from real data, it is possible to induce a herding effect toward a new community with the concerted effort of a small number of users.

REFERENCES

Ashton Anderson, Daniel Huttenlocher, Jon Kleinberg, and Jure Leskovec. 2013. Steering User Behavior with Badges. In *Proceedings of the Twenty-Second International Conference on World Wide Web (WWW '13)*. International World Wide Web Conferences Steering Committee, Republic and Canton of Geneva, Switzerland, 95–106. http://dl.acm.org/citation.cfm?id=2488388.2488398

Ido Erev and Alvin E Roth. 1998. Predicting How People Play Games: Reinforcement Learning in Experimental Games with Unique, Mixed Strategy Equilibria. *American Economic Review* 88, 4 (September 1998), 848–81.

Kevin Morris. 2012. How Reddit's cofounders built Reddit with an army of fake accounts. *The Daily Dot* (June 2012). http://www.dailydot.com/business/steve-huffman-built-reddit-fake-accounts/

Andrew Y. Ng and Stuart J. Russell. 2000. Algorithms for Inverse Reinforcement Learning. In *Proceedings of the Seventeenth International Conference on Machine Learning (ICML '00)*. Morgan Kaufmann Publishers Inc., San Francisco, CA, USA, 663–670. http://dl.acm.org/citation.cfm?id=645529.657801

Yee Whye Teh, Michael I. Jordan, Matthew J. Beal, and David M. Blei. 2006. Hierarchical Dirichlet Processes. *Journal of the American Statistical Association* 101, 476 (2006), pp. 1566–1581.

Fang Wu, D.M. Wilkinson, and B.A. Huberman. 2009. Feedback Loops of Attention in Peer Production. In *International Conference on Computational Science and Engineering, 2009. CSE '09.*, Vol. 4. 409–415. DOI:http://dx.doi.org/10.1109/CSE.2009.430

Haiyi Zhu, Amy Zhang, Jiping He, Robert E. Kraut, and Aniket Kittur. 2013. Effects of Peer Feedback on Contribution: A Field Experiment in Wikipedia. In *Proceedings of the SIGCHI Conference on Human Factors in Computing Systems (CHI '13)*. ACM, New York, NY, USA, 2253–2262. DOI:http://dx.doi.org/10.1145/2470654.2481311